\documentclass[prd,twocolumn,showpacs,floatfix]{revtex4}
\usepackage{hyperref,amssymb,amsmath,mathrsfs,bm,graphicx}
\usepackage{enumitem}
\usepackage{ulem}
\usepackage{color}
\begin{document}
\title{Conformally flat  polytropes for anisotropic matter}
\author{L. Herrera}
\email{lherrera@usal.es}
\affiliation{Escuela de F\'\i sica, Facultad de Ciencias, Universidad Central de Venezuela, Caracas 1050, Venezuela}
\author{W. Barreto}
\email{wbarreto@ula.ve}
\affiliation{Centro de F\'\i sica Fundamental, Facultad de Ciencias,
Universidad de Los Andes, M\'erida 5101, Venezuela}
\author{A. Di Prisco}
\email{adiprisc@ciens.ucv.ve}
\affiliation{Escuela de F\'\i sica, Facultad de Ciencias, Universidad Central de Venezuela, Caracas 1050, Venezuela}
\author{J. Ospino}
\email{j.ospino@usal.es}
\affiliation{Departamento de Matem\'atica Aplicada and Instituto Universitario de F\'isica
Fundamental y Matematicas,, Universidad de Salamanca,
Salamanca  37007, Spain}

\begin{abstract}
We analyze in detail conformally flat spherically symmetric fluid distributions, satisfying a polytropic equation of state.  Among the two possible families of relativistic polytropes, only one contains models which satisfy all the required physical conditions. The ensuing configurations are necessarily anisotropic and  show interesting physical properties. Prospective applications of the presented models to the study of super-Chandrasekhar white dwarfs, are discussed.
\end{abstract}
\date{\today}
\maketitle

\section{Introduction}
In two recent papers the general formalism to study  polytropes for anisotropic matter has been reported, both in the Newtonian  \cite{1p} and in the general relativistic regimes  \cite{2p}. The motivations to undertake such a task were exposed in detail in \cite{2p} accordingly we shall not repeat them here.

The theory of polytropes is based on the polytropic equation of state, which in the Newtonian case reads
\begin{equation}P=K\rho_0^{\gamma}=K\rho_0^{1+1/n} ,\label{Pol}\end{equation}
where $P$ and $\rho_0$ denote the isotropic pressure and the  mass (baryonic) density,  respectively. Constants $K$, $\gamma$, and $n$ are usually called  the polytropic constant, polytropic exponent, and polytropic index, respectively.

In the general relativistic anisotropic case, two  extensions of the above equation of state are possible, namely:

\begin{enumerate}
\item \begin{equation}P_r=K\rho_0^{\gamma}=K\rho_0^{1+1/n} \label{p1},\end{equation} 
i.e. the original polytropic equation of state (\ref{Pol}) is conserved, and from simple thermodynamic considerations it follows  that the total energy density $\rho$ and the baryonic (mass density) $\rho_0$ are related through the equation (see \cite{2p} for details)
\begin{equation}
\rho=\rho_0+nP_r,
\label{nueva1}
\end{equation}
where $P_r$   denotes the radial pressure (see below).
\item \begin{equation}P_r=K\rho^{\gamma}=K\rho^{1+1/n} \label{p2}.\end{equation} 

 In this case the baryonic density $\rho_0$ is replaced by the total energy density $\rho$ in the polytropic equation of state. Also, it can be shown that  the relationship between the two densities is given by (see \cite{2p} for details)
\begin{equation}
\rho=\frac{\rho_0}{\left(1-K\rho^{1/n}_0\right)^n}.
\label{nueva2}
\end{equation}
\end{enumerate}
As it should be expected, the assumption of either (\ref{p1}) or (\ref{p2}) is not enough to integrate completely the  field equations, since the appearance of  two principal stresses (instead of one as in the isotropic case) leads to  a system of two equations for three unknown functions.

Thus  in  order to integrate the obtained system of equations we need to provide further information about the anisotropy inherent to the problem under consideration. For doing that, in Ref. \cite{2p}   the particular  ansatz  of Ref. \cite{anis4} has been assumed, which  allows  specific modelling. This  method  links the obtained  models continuously with the isotropic case.

Here we shall proceed differently. In order to integrate our equations we shall assume the vanishing of the Weyl tensor (conformally flat condition). The motivation for this assumption is based on the role of the Weyl tensor in the structure and evolution of self-gravitating systems. Indeed,  for spherically symmetric distributions of fluid, the Weyl tensor may be defined exclusively in terms of the density contrast and the local anisotropy  of the pressure (see Sec. III), which in turn are known to affect the fate of gravitational collapse (see Refs. \cite{large2, large3, large4, large5, pla98, large1, large9, large10, large11, nueva12} and references therein). Thus, the vanishing of the Weyl tensor establishes a specific relationship between density contrast and local anisotropy of pressure.

Furthermore, the conformally flat condition implies energy density homogeneity for the perfect (isotropic pressure) fluid sphere. This in turn implies that there are not bounded isotropic (in pressure) conformally flat polytropes. Accordingly, our models  are necessarily anisotropic (in pressure), and are not continuously linked to the isotropic sphere. The specific anisotropy, resulting from the two basic assumptions (conformal flatness and polytropic equation of state), is related to the density contrast,  and depends only on the two parameters  characterizying each model.

In the next section we shall briefly review the main aspects of the anisotropic polytropes. Next, in Sec. III, we incorporate the conformally flat condition into our formalism. Sec. IV is devoted to the analysis of the obtained models. For these models we shall also calculate the Tolman mass, whose behaviour allows to characterize them in more detail.

Finally, we shall conclude with a summary, and  some  possible extensions, of our results.

\section{The general relativistic polytrope for anisotropic fluid}
We consider spherically symmetric static distributions of anisotropic fluid (principal stresses unequal), bounded by a spherical surface $\Sigma$, defined by the equation $r=r_\Sigma=$ const. \noindent

 The line element is given in Schwarzschild-like coordinates by 
\begin{equation}ds^2=e^{\nu} dt^2 - e^{\lambda} dr^2 -r^2 \left( d\theta^2 + \sin^2\theta d\phi^2 \right),\label{metric}\end{equation}\noindent 
where $\nu$ and $\lambda$ are functions of $r$.
We number the coordinates: $x^0=t; \, x^1=r; \, x^2=\theta; \, x^3=\phi$. We use geometric units and therefore we have $c=G=1$.

If we allow the principal stresses to be unequal, then the energy momentum  tensor in the canonical form reads
\begin{equation}
T^{\mu}_{\nu}=\rho u^{\mu}u_{\nu}-  P
h^{\mu}_{\nu}+\Pi ^{\mu}_{\nu},
\label{24'}
\end{equation}
where $\rho$ is the energy density,  $P$ is  the isotropic pressure, and $\Pi_{\mu \nu}$ the anisotropic pressure tensor,
with 
\begin{equation}
h^{\mu}_{\nu}=\delta^{\mu}_{\nu}-u^{\mu}u_{\nu},\quad \Pi
^{\mu}_{\nu}=\Pi\left(s^{\mu}s_{\nu}+\frac{1}{3}h^{\mu}_{\nu}\right),
\label{nt}
\end{equation}
where
\begin{equation}
u^\mu=\left(e^{-\nu/2}, \, 0,\, 0, \, 0\right),\label{umu}
\end{equation}
denotes the four velocity of the fluid,
and  $s^\mu$ is defined as
\begin{equation}
s^{\mu}=(0, e^{-\lambda/2},0,0),
\end{equation}
satisfying
$s^{\mu}u_{\mu}=0$,
$s^{\mu}s_{\mu}=-1$.

For our purposes in this work, it would be more convenient to introduce the following  two auxiliary variables, $P_r$ and $P_\perp$, hereafter  referred to as the radial and tangential pressures, respectively:  

\begin{equation}
P_r=s^\alpha s^\beta T_{\alpha \beta},\qquad P_\perp=k^\alpha k^\beta T_{\alpha \beta},
\end{equation}
where $k^\alpha$ is a  unit spacelike vector (orthogonal to $u^\alpha$ and $s^\alpha$).

In terms of the above variables, we have
\begin{equation}
\Pi=P_r-P_\perp ; \qquad P=\frac{P_r+2 P_\perp}{3},
\label{ns}
\end{equation}
from where the physical meaning of $P_r$ and $P_\perp$ becomes evident, and the energy momentum can be written under the form

\begin{equation}
T_{\mu\nu} = \left(\rho+P_\bot\right)u_\mu u_\nu - P_\bot g_{\mu\nu} + \left(P_r-P_\bot\right)s_\mu s_\nu. 
\end{equation}

 The metric (\ref{metric}) has to satisfy Einstein field equations which in our case read \cite{2p}:
\begin{equation}\rho=-\frac{1}{8 \pi}\left[-\frac{1}{r^2}+e^{-\lambda}\left(\frac{1}{r^2}-\frac{\lambda'}{r} \right)\right],\label{fieq00}\end{equation}
\begin{equation}P_{r} =-\frac{1}{8\pi}\left[\frac{1}{r^2} - e^{-\lambda}\left(\frac{1}{r^2}+\frac{\nu'}{r}\right)\right],\label{fieq11}\end{equation}
\begin{eqnarray}P_{\perp} = \frac{1}{8 \pi}\left[ \frac{e^{-\lambda}}{4}\left(2\nu''+\nu'^2 -\lambda'\nu' + 2\frac{\nu' - \lambda'}{r}\right)\right],\label{fieq2233}\end{eqnarray}
where prime denotes derivative with respect to $r$.

At the outside of the fluid distribution, the spacetime is that of Schwarzschild, given by
\begin{eqnarray}
ds^2&=& \left(1-\frac{2M}{r}\right) dt^2 - \left(1-\frac{2M}{r}\right)^{-1}d{r}^2 \nonumber \\
&&\;\;\;\;\;\;\;\;\;\;\;\;\;\;\;\;\;\;\;\;\;\;
-{r^2} \left(d\theta^2 + \sin^2\theta d\phi^2 \right),\label{Schw_ext}
\end{eqnarray}
In order to match smoothly the two metrics above on the boundary surface $r=r_\Sigma$, we must require the continuity of the first and the second fundamental forms across that surface (Darmois conditions). Then it follows 
\begin{equation}
e^{\nu_\Sigma}=1-\frac{2M}{r_\Sigma},\label{enusigma}
\end{equation}
\begin{equation}
e^{-\lambda_\Sigma}=1-\frac{2M}{r_\Sigma},\label{elambdasigma}
\end{equation}
\begin{equation}P_{r\Sigma}=0,\label{PQ}
\end{equation}
where the subscript $\Sigma$ indicates that the quantity is evaluated at the boundary surface $\Sigma$.\noindent

From the previous  expressions it is a simple matter to prove that the hydrostatic equilibrium  equation now reads
\begin{equation}
P'_r=-\frac{\;\nu'}{2}\left(\rho+P_r\right)+\frac{2\left(P_\bot-P_r\right)}{r}.\label{Prp}
\end{equation}
This is the  generalized Tolman-Opphenheimer-Volkoff equation for anisotropic matter. 
Alternatively, using 
\begin{equation}
\frac{\;\nu'}{2} = \frac{m + 4 \pi P_r r^3}{r \left(r - 2m\right)},
\label{nuprii}
\end{equation}
we may write
\begin{equation}
P'_r=-\frac{(m + 4 \pi P_r r^3)}{r \left(r - 2m\right)}\left(\rho+P_r\right)+\frac{2\left(P_\bot-P_r\right)}{r},\label{ntov}
\end{equation}
 where   the mass function $m(r)$, as usually, is  defined by 
\begin{equation}
e^{-\lambda}=1-2m/r, \qquad m(r) = 4 \pi \int^r_0{\rho r^2 dr}.\label{mass}
\end{equation}
We shall consider the  two  cases defined by equations (\ref{p1}) and (\ref{p2}),  to extend the polytropic equation of state to anisotropic matter. In order to close the system of resulting equations, we shall further assume the vanishing of the Weyl tensor.

All the models   have to satisfy physical requirements such as:
\begin{equation}
\rho >0, \qquad  \frac{P_r}{\rho}\leq 1, \qquad  \frac{P_\perp }{\rho} \leq 1.
\label{conditions}
\end{equation}

We shall next proceed to describe briefly each case (see Ref.  \cite{2p}) for details).

\subsection{Case I}
Assuming Eq. (\ref{p1}), let us introduce  the following variables
\begin{equation}
\alpha=P_{rc}/\rho_{c},\quad r=\xi/A,  \quad A^2=4 \pi \rho_{c}/\alpha (n+1)\label{alfa},\end{equation}

\begin{equation}\psi_{0}^n=\rho_{0}/\rho_{0c},\quad v(\xi)=m(r) A^3/(4 \pi\rho_{c}),\label{psi}\end{equation}
where subscript $c$ indicates that the quantity is evaluated at the center. At the boundary surface $r=r_\Sigma$ ($\xi=\xi_\Sigma$) we have $\psi_0(\xi_\Sigma)=0$. 

Then, the generalized Tolman-Opphenheimer-Volkoff equation becomes
\begin{widetext}
\begin{eqnarray}
\xi^2 \frac{d\psi_{0}}{d\xi}\left[\frac{1-2(n+1)\alpha v/\xi}{(1-n\alpha)+(n+1)\alpha \psi_{0}}\right]+v+\alpha\xi^3 \psi_{0}^{n+1}+\frac{2\Delta \psi_0^{-n}\xi}{P_{rc}(n+1)} \left[\frac{1-2\alpha (n+1)v/\xi}{(1-n\alpha)+(n+1)\alpha \psi_0}\right]=0,\label{TOV1anis_WB}
\end{eqnarray}
\end{widetext}
where $\Delta=-\Pi=P_\bot-P_r$.

On the other hand we obtain from the mass function definition (\ref{mass})  and Eq. (\ref{fieq00}),
\begin{equation}
m'=4 \pi r^2 \rho\label{mprima}\end{equation}or
\begin{equation}
\frac{dv}{d\xi}=\xi^2 \psi_{0}^n (1-n\alpha+n\alpha\psi_{0}).\label{veprima}\end{equation}

In this case, conditions (\ref{conditions}) read
\begin{eqnarray}
n\alpha<1,\qquad  \frac{\alpha \psi_0}{1-n\alpha+n\alpha \psi_0}\leq 1, \nonumber \\ \frac{3v/\xi^3+\alpha \psi_0^{n+1}}{\psi_0^n(1-n\alpha+n\alpha\psi_o)}-1\leq1.
\label{conditionsII}
\end{eqnarray}

Combining Eqs. (\ref{TOV1anis_WB}) and (\ref{veprima}) we are led to the generalized Lane-Emden equation for this case (see Ref. \cite{2p}).

\subsection{Case II}

In this case the assumed equation of state is (\ref{p2}), then, introducing
\begin{equation}
\psi^n=\rho/\rho_{c},\label{psi2}\end{equation}
the generalized Tolman-Opphenheimer-Volkoff equation becomes

\begin{widetext}

\begin{eqnarray}
\xi^2 \frac{d\psi}{d\xi}\left[\frac{1-2(n+1)\alpha v/\xi}{1+\alpha \psi}\right]+v+\alpha\xi^3 \psi^{n+1}+\frac{2\Delta \psi^{-n}\xi}{P_{rc}(n+1)} \left[\frac{1-2\alpha(n+1)v/\xi}{1+\alpha \psi}\right]=0,\label{TOV2anis_WB}
\end{eqnarray}
\end{widetext}
and from {Eq.} (\ref{mprima})
\begin{equation}\frac{dv}{d\xi}=\xi^2 \psi^n.\label{veprima2}\end{equation}
In this case, conditions (\ref{conditions}) read:
\begin{equation}
\rho>0, \qquad \alpha \psi \leq 1, \qquad \frac{3v}{\xi^3 \psi^n}+\alpha \psi-1 \leq1.
\label{conditionsIII}
\end{equation}
Once again, the combination of Eqs. (\ref{TOV2anis_WB}) and (\ref{veprima2}) leads to the generalized Lane-Emden equation for this case (see Ref. \cite{2p} for details).

Equations (\ref{TOV1anis_WB}), (\ref{veprima}) or (\ref{TOV2anis_WB}), (\ref{veprima2}), form a system of two first order  ordinary differential equations for the three unknown functions: $\psi (\psi_0), v, \Delta$, depending on a duplet of parameters $n, \alpha$. Thus, it is obvious that in order to proceed further with the modeling of a compact object, we need to provide additional information. Such information, of course, depends on the specific physical problem under consideration. For the reasons exposed in the Introduction, here we shall assume the conformally flat condition (vanishing of the Weyl tensor).
\begin{figure}
\includegraphics[width=3.in,height=1.5in,angle=0]{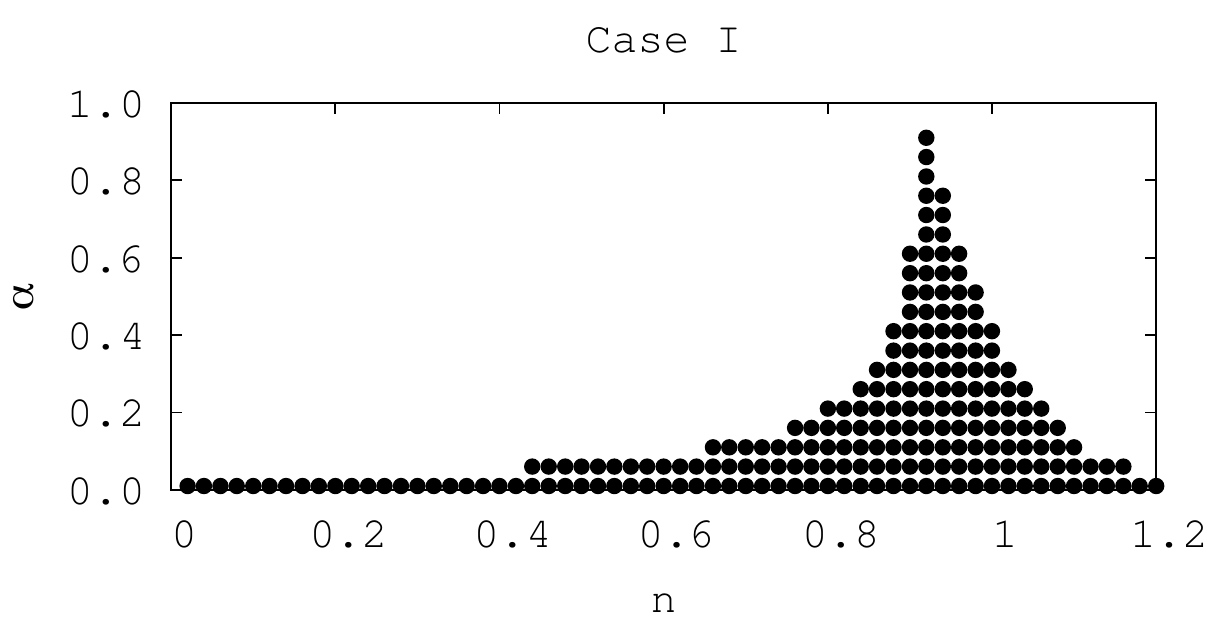}
\caption{Case I: Each point represents a duplet ($n$,$\alpha$ [multiplied by $10^{8}$]) for which the conditions (\ref{conditionsII}) are satisfied in the whole distribution.}
\label{fig:figure1}
\end{figure}
\section{The Weyl tensor and the conformally flat condition}
In the spherically symmetric case, it can be shown that all the non-vanishing  components of the Weyl tensor $C_{\alpha \gamma \beta \delta}$ can be expressed through the component $C^3_{232}$ (see Ref. \cite{HS} for details).
Thus, the Weyl tensor is described by the single function
\begin{widetext}
\begin{equation}
W \equiv \frac{r}{2} C^3_{232} =
\frac{r^3 e^{-\lambda}}{6} \left(
\frac{e^\lambda}{r^2} + \frac{\nu' \lambda'}{4} -
\frac{1}{r^2} - \frac{\nu'^2}{4} - \frac{\nu''}{2} -
\frac{\lambda' - \nu'}{2r}
\right).
\label{W}
\end{equation}
\end{widetext}
Then, the  following relation may be established (see for example Ref. \cite{HS})
\begin{equation}
W = - \frac{4}{3} \pi \int_0^r{r^3 \rho' dr} +
\frac{4}{3} \pi r^3 \left(P_r- P_{\perp}\right).
\label{Wint}
\end{equation}
The above equation expresses the Weyl tensor in terms of the energy density contrast and the local anisotropy of pressure.

It has been shown in Ref. \cite{arrow7} that the conformally flat condition ($W=0$) can be integrated, producing (see Refs. \cite{arrow7} for details)
\begin{equation}
e^\nu = \tilde{B}^2 r^2 \cosh^2{\left[\int{\frac{e^{\lambda/2}}{r}dr}+ \tilde{C}\right]}
\label{fint}
\end{equation}
where $\tilde{B}$ and $\tilde{C}$ are  constants of integration. 

Thus the conformally flat condition reduces  the number of unknown functions, which in turn allows us to integrate either Eqs. (\ref{TOV1anis_WB})-(\ref{veprima}) or Eqs. (\ref{TOV2anis_WB})-(\ref{veprima2}).

However, instead of  using Eq. (\ref{fint}), we shall proceed differently.

 First,  we observe that from Eqs. (\ref{fieq11}), (\ref{fieq2233}) and
$W=0$, it follows that:
\begin{equation}
\Delta=\frac{r}{8\pi}\left(\frac{e^{-\lambda}-1}{r^2}\right)^\prime,
\label{Pilambda}
\end{equation}
producing for the case I,

\begin{equation}
\Delta=\rho_c\left[\frac{3 v}{\xi^3}-\psi_0^n(1-n\alpha+n\alpha \psi_0 )\right],
\label{Pilambdabis}
\end{equation}
where Eqs. (\ref{alfa}) and (\ref{veprima}) have been used.

Feeding back this last expression into Eq. (\ref{TOV1anis_WB}), we obtain
\begin{widetext}
\begin{eqnarray}
\xi^2 \frac{d\psi_{0}}{d\xi}\left[\frac{1-2(n+1)\alpha v/\xi}{(1-n\alpha)+(n+1)\alpha \psi_{0}}\right]+v+\alpha\xi^3 \psi_{0}^{n+1}\nonumber \\+\frac{2\psi_0^{-n}\xi}{\alpha(n+1)} \left[\frac{1-2\alpha (n+1)v/\xi}{(1-n\alpha)+(n+1)\alpha \psi_0}\right]\left[\frac{3 v}{\xi^3}-\psi_0^n(1-n\alpha+n\alpha \psi_0 )\right]=0.\label{TOV1anis_WBbis}
\end{eqnarray}
\end{widetext}

Thus for the case I, the two equations describing the conformally flat polytrope are (\ref{veprima}) and (\ref{TOV1anis_WBbis}), which is a system of two equations for two unknown functions, and can be solved for any set of the parameters $n,\alpha$.

For the case II we obtain
\begin{equation}
\Delta= \rho_c\left(\frac{3 v}{\xi^3}-\psi^n\right),
\label{PilambdabisII}
\end{equation}
producing
\begin{widetext}
\begin{eqnarray}
\xi^2 \frac{d\psi}{d\xi}\left[\frac{1-2(n+1)\alpha v/\xi}{1+\alpha \psi}\right]+v+\alpha\xi^3 \psi^{n+1}+\frac{2 \psi^{-n}\xi}{\alpha(n+1)} \left[\frac{1-2\alpha(n+1)v/\xi}{1+\alpha \psi}\right](\frac{3 v}{\xi^3}-\psi^n)=0.\label{TOV2anis_WBbis}
\end{eqnarray}
\end{widetext}

Thus in this latter case, the two equations fully describing the polytrope are (\ref{veprima2}) and (\ref{TOV2anis_WBbis}).

It will  be useful to  calculate the Tolman mass, which is   a measure of the active gravitational mass (see  Refs. \cite{ pla98, TM} and references therein), and which may be written as

\begin{equation}
m_T=e^{(\nu+\lambda)/2} (m+4\pi P_r r^3).
\label{mt1}
\end{equation}

Alternatively, the following expression can be found for the Tolman mass (see Eq. (32) in Ref. \cite{HS})
\begin{eqnarray}
m_{T} &=& M \left(\frac{r}{r_\Sigma}\right)^3  \nonumber \\&+& r^3 
\int^{r_\Sigma}_{r}{e^{\left(\nu + \lambda\right)/2} 
\left[\frac{3}{\tilde r^4} W - \frac{4 \pi \Delta}{\tilde r} 
\right]d \tilde r}.
\label{exbu}
\end{eqnarray}

The functions  $\lambda$, $m$ and $P_r$ in the above expressions, are obtained directly by integration of Eqs. (\ref{TOV1anis_WBbis}) and (\ref{veprima}) for the case I, and  Eqs. (\ref{veprima2}) and (\ref{TOV2anis_WBbis}), for the  case II.  Thus we only need an expression for $\nu$ which can be obtained  directly from Eq. (\ref{fint}). However it is easier to obtain $\nu$ directly from the integration of Eq. (\ref{nuprii}).  

Thus we have, 
\begin{equation}
\nu=\nu_\Sigma -2\int^{r_\Sigma}_r\frac{(m+4\pi P_r r^3)}{r(r-2m)}dr.
\label{mt11}
\end{equation}

In order to see how the Tolman mass distributes through the sphere in the process of  contraction (slow and adiabatic), it would be convenient to introduce the following dimensionless variables:
\begin{equation}
x=\frac{r}{r_\Sigma}=\frac{\xi}{\tilde A},\qquad y=\frac{M}{r_\Sigma},\qquad \tilde m=\frac{m}{M},\qquad \tilde A=r_\Sigma A.
\label{nm1}
\end{equation}

Then, using Eq. (\ref{exbu}) we find for the the cases I and II, respectively:
\begin{widetext}
\begin{equation}
\frac{m_T}{M}=x^3-\frac{\alpha(n+1)x^3\tilde A^2}{y}\int^{1}_x \left\{\sqrt{\frac{1-2y}{1-2v\alpha(n+1)/ \tilde A x}} exp \left \{-\frac{\alpha(n+1)}{\tilde A}\int^{1}_x\frac{v+\tilde A^3 x^3\alpha \psi^{n+1}_0}{ x^2\left[1-2v\alpha(n+1)/\tilde A x \right]}dx\right\}\frac{\Omega dx}{x}\right\},
\label{nmIbis}
\end{equation}
\end{widetext}
\begin{widetext}
\begin{equation}
\frac{m_T}{M}=x^3-\frac{\alpha(n+1)x^3\tilde A^2}{y}\int^{1}_x \left\{\sqrt{\frac{1-2y}{1-2v\alpha(n+1)/ \tilde A x}} exp \left \{-\frac{\alpha(n+1)}{\tilde A}\int^{1}_x\frac{v+\tilde A^3 x^3\alpha \psi^{n+1}}{ x^2\left[1-2v\alpha(n+1)/\tilde A x \right]}dx\right\}\frac{\Omega dx}{x}\right\},
\label{nmIIbis}
\end{equation}
\end{widetext}
where the fact that 
\begin{equation}
y=\alpha(n+1) \frac{v_\Sigma}{\xi_\Sigma},
\label{nmIV}
\end{equation}
has been used,
and we have introduced   the quantity $\Omega$ defined by 
\begin{equation}
\Omega\equiv \frac{\Delta}{\rho_c}.
\label{rela1}
\end{equation}
Using Eqs. (\ref{Pilambdabis}) and  (\ref{PilambdabisII}), we obtain for the cases I and II, respectively:
\begin{equation}
\Omega=\left[\frac{3v}{\xi^3}-\psi^n_0(1-n\alpha+n\alpha \psi_0) \right],
\label{rela2}
\end{equation}
and
\begin{equation}
\Omega=\left(\frac{3v}{\xi^3}-\psi^n \right).
\label{rela3}
\end{equation}

The full set of equations deployed above  has been integrated for both cases (I and II), and a wide range of values of different parameters ($n,\alpha$). In what follows we analyze the most relevant results emerging from a selection of  the whole set of obtained models.
\begin{figure}
\includegraphics[width=3.in,height=3.in,angle=0]{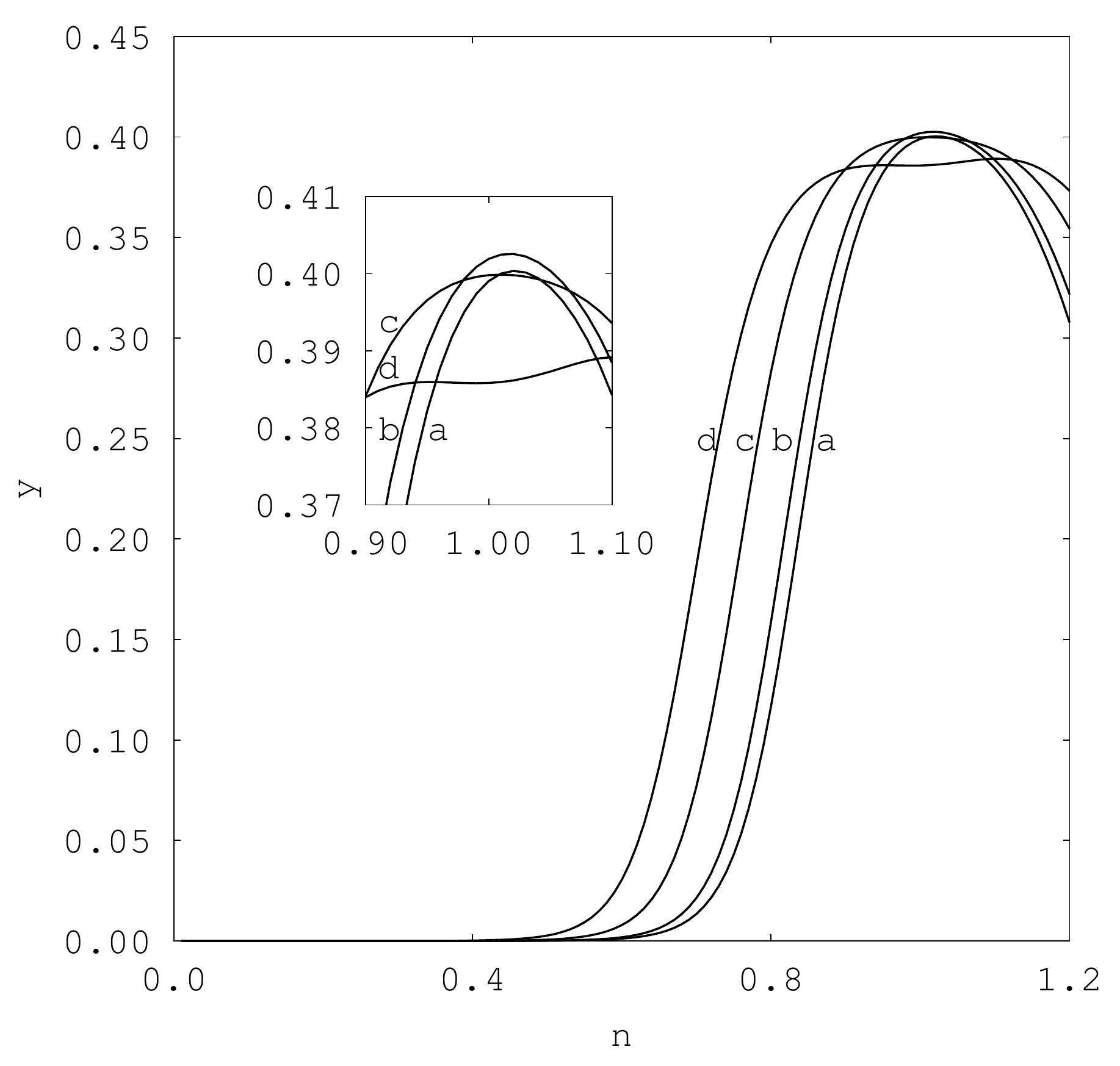}
\caption{Case I: $y$ as a function of $n$ for $\alpha$: $8\times 10^{-11}$  (curve a), $10^{-10}$ (curve b), $2\times 10^{-10}$ (curve c), $4\times 10^{-10}$ (curve d). The inserted graph shows the same function in the interval $n \in [0.9, 1.1]$.}
\label{fig:figure2}
\end{figure}
\begin{figure}
\includegraphics[width=3.in,height=3.in,angle=0]{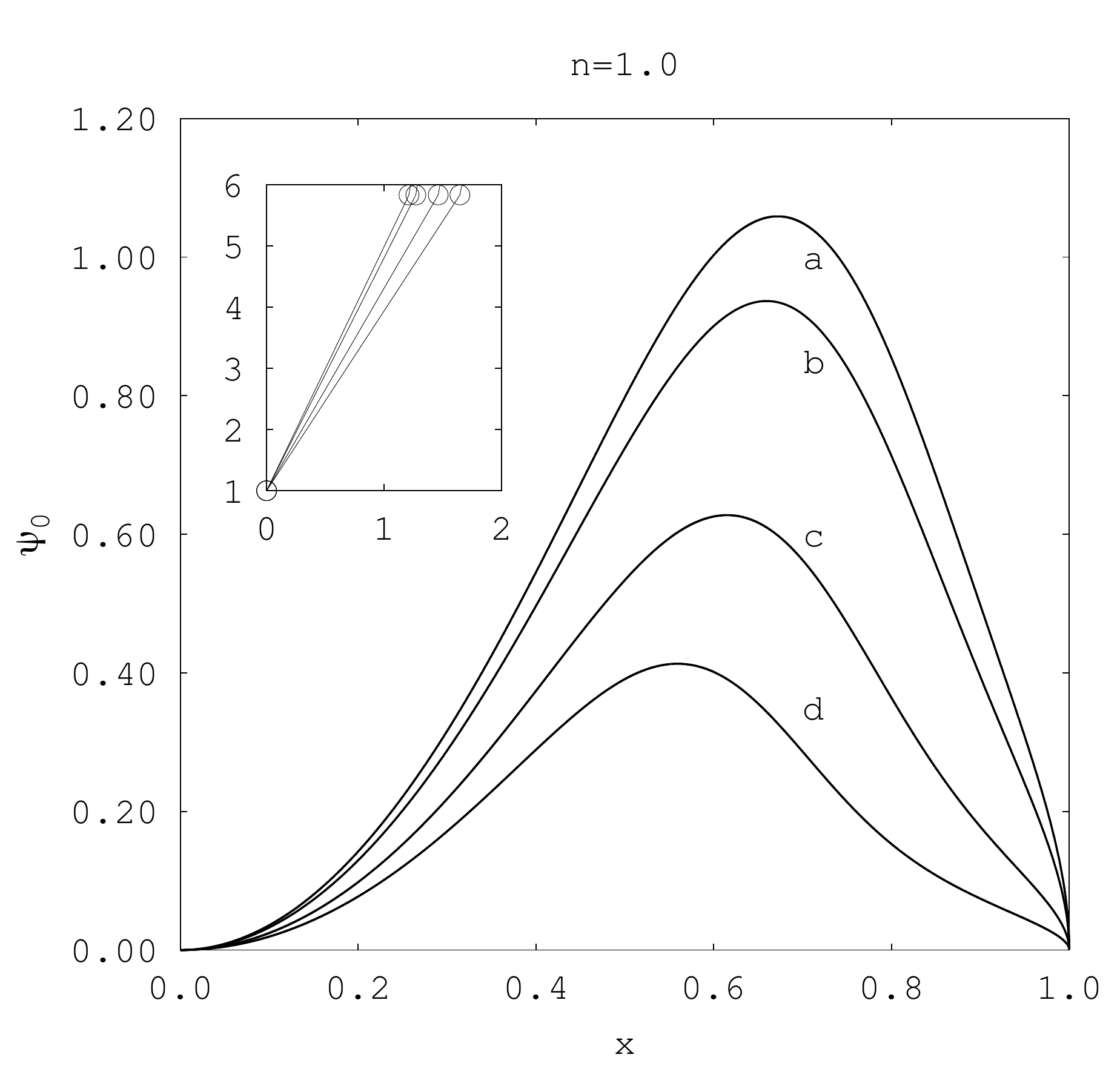}
\caption{Case I: $\psi_0$ (multiplied by $10^{-10}$) as a function of $x$ for $n=1$ and $\alpha (y)$: $8\times 10^{-11} (0.3991)$  (curve a), $10^{-10} (0.4019)$ (curve b), $2\times 10^{-10} (0.3998)$ (curve c), $4\times 10^{-10} (0.3858)$ (curve d). The inserted graph shows the first two points of $\psi_0$ (not multiplied by $10^{-10}$) as a function of $x$ (multiplied by $10^5$). Curves from left to right correspond to curves $a$ $\rightarrow$ $d$. The parameter choice is qualitatively representative of all models.}
\label{fig:figure3}
\end{figure}
\begin{figure}
\includegraphics[width=3.in,height=3.in,angle=0]{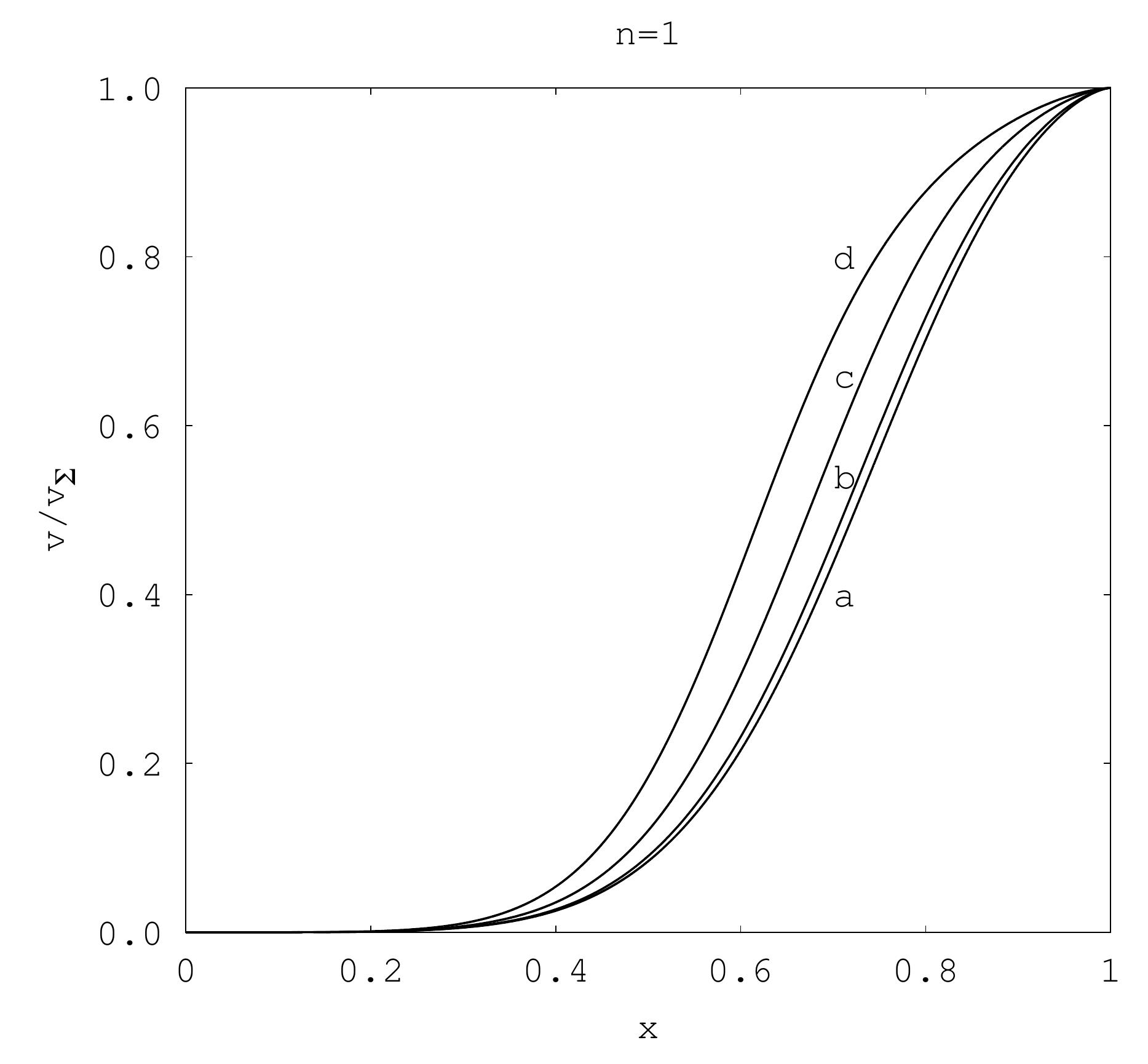}
\caption{Case I: $v/v_\Sigma$ as a function of $x$ for $n=1$ and $\alpha (y)$: $8\times 10^{-11} (0.3991)$  (curve a), $10^{-10} (0.4019)$ (curve b), $2\times 10^{-10} (0.3998)$ (curve c), $4\times 10^{-10} (0.3858)$ (curve d).}
\label{fig:figure4}
\end{figure}
\begin{figure}
\includegraphics[width=3.in,height=3.in,angle=0]{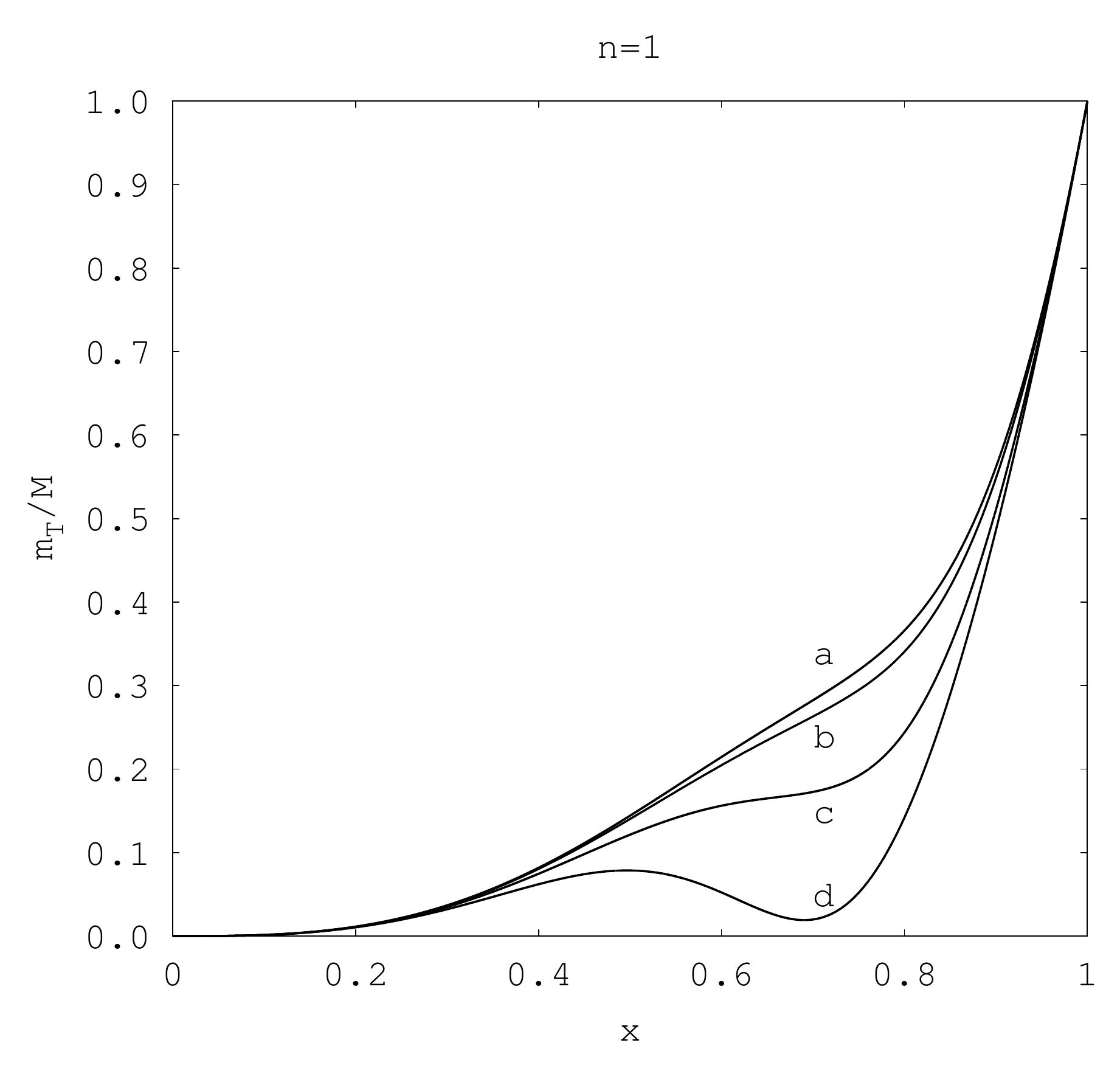}
\caption{Case I: $m_T/M$ as a function of $x$ for $n=1$ and $\alpha (y)$: $8\times 10^{-11} (0.3991)$  (curve a), $10^{-10} (0.4019)$ (curve b), $2\times 10^{-10} (0.3998)$ (curve c), $4\times 10^{-10} (0.3858)$ (curve d).}
\label{fig:figure5}
\end{figure}
\begin{figure}
\includegraphics[width=3.in,height=3.in,angle=0]{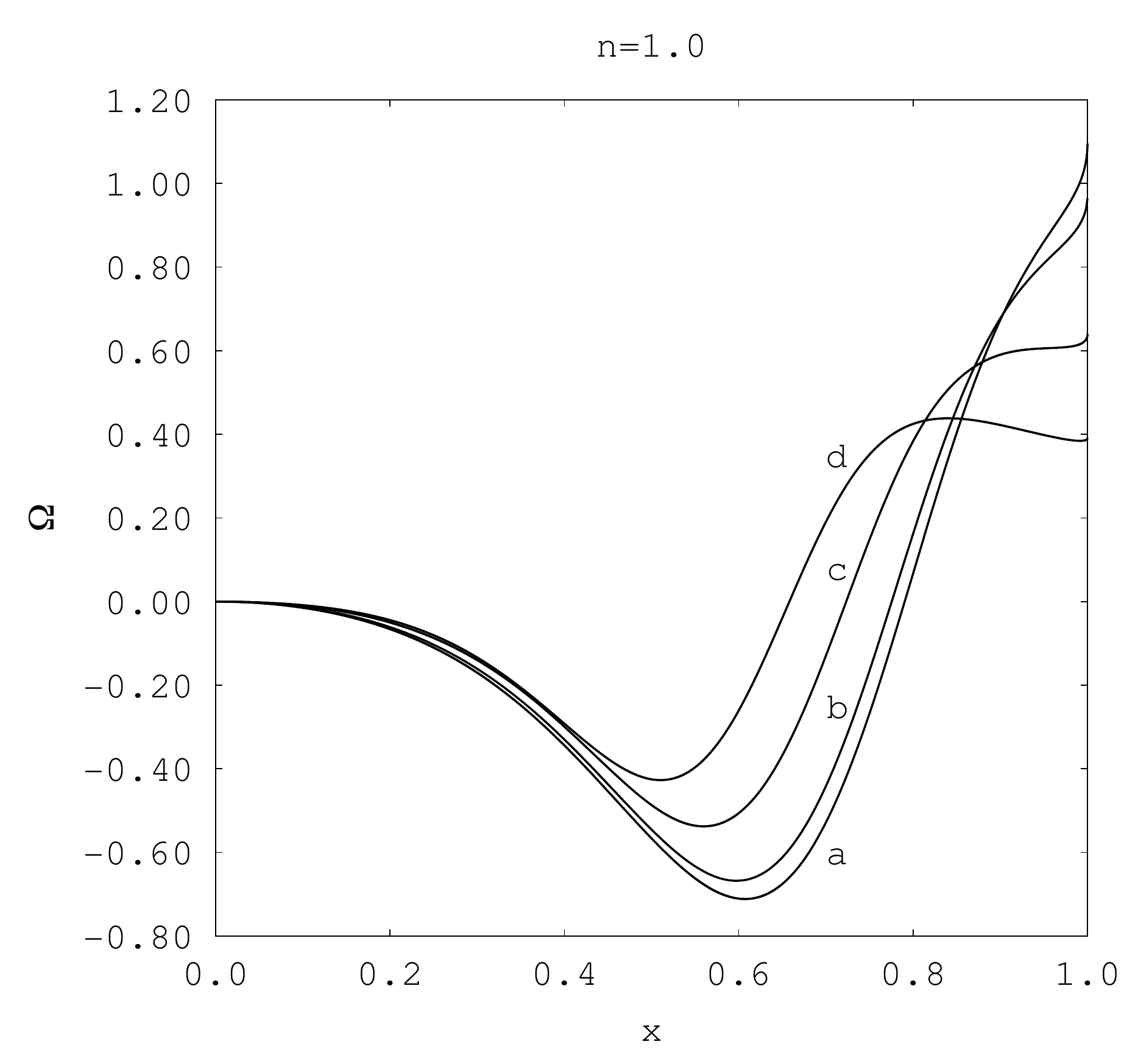}
\caption{Case I: $\Omega$ (multiplied by $10^{-10}$) as a function of $x$ for $n=1$ and $\alpha (y)$: $8\times 10^{-11} (0.3991)$  (curve a), $10^{-10} (0.4019)$ (curve b), $2\times 10^{-10} (0.3998)$ (curve c), $4\times 10^{-10} (0.3858)$ (curve d).}
\label{fig:figure6}
\end{figure}
\section{Results}
Using the boundary conditions
\begin{equation}
v(0)=0,
\end{equation}
for both cases, and
\begin{equation}
\psi_0(0)=1,\; \psi_0(\xi_{\Sigma})=0,\;\; \; \psi(0)=1,\; \psi(\xi_\Sigma)=0.
\end{equation}
depending on the respective case.

We have integrated numerically Eqs. (\ref{veprima}), (\ref{TOV1anis_WBbis}) and (\ref{veprima2}), (\ref{TOV2anis_WBbis}), using the four order Runge-Kutta method. We stopped the integration once the surface $\xi_\Sigma$ is reached. Near $\xi=0$ we set the asymptotic approximation $d\psi_0/d\xi=0$ and $d\psi/d\xi=0$, respectively.
To calculate integrals in Eqs. (\ref{nmIbis}) and (\ref{nmIIbis}) we used a simple second order midpoint recursive formula.

\subsection{Case I}
Figure \ref{fig:figure1} depicts the region of allowed solutions (those satisfying all the requirements (\ref{conditionsII})). We  observe that physically admissible solutions exist only for very small values of $\alpha$, meaning that the fluid around the centre is far from the relativistic regime. Notwithstanding, as shown in Fig. \ref{fig:figure2}, for  a range of values of $n$, the resulting  configurations may be quite compact, approaching in some cases the value $y=0.41$, close to  the upper limit for isotropic spheres ($y=0.44$). It is worth recalling that the local anisotropy of pressure has a direct impact on the maximal value of the surface potential (see Refs. \cite{mx1, mx2, mx3, mx5, mx6, mx4, mx7, F, BL, HRW} and references therein).

Figures \ref{fig:figure3} and \ref{fig:figure4} show the integration of Eqs. (\ref{TOV1anis_WBbis}) and (\ref{veprima}), for the indicated values of the duplet $n, \alpha$. However the  behaviour exhibited in both figures, is qualitatively representative for any other values of the parameters $n,\alpha$ (among those that produce physically admissible models).

  We observe that while the mass function is a monotonically increasing function of the radial coordinate (as expected), the spatial derivative of $\psi_0$  changes of sign within  the sphere. This last behaviour, of course, is possible  by the anisotropy of the fluid, and cannot be present in locally isotropic models. 

The behaviour of the Tolman mass is also very peculiar. Indeed, looking at Figure \ref{fig:figure2}, we see that for the models depicted in Figure \ref{fig:figure5}, the slow (adiabatic) contraction of the system is described by the sequence $d\rightarrow a \rightarrow c \rightarrow b$, corresponding  to the evolution from smaller to larger values of $y$. Obviously, more stable configurations support larger surface gravitational potentials (larger $y$). Therefore,   this  sequence  corresponds to the evolution  from the less stable (smaller $y$), to the more stable (larger $y$) object. Let us now try to understand such a behaviour at the light of Figure \ref{fig:figure5}

First of all, we observe that for the cases $d$ and $c$, there is a sharper migration of the Tolman mass towards the boundary surface, than the one exhibited by $a$ and $b$. Such  an effect  might suggest, due to the physical interpretation of the Tolman mass as a measure of the active gravitational mass,   more stability of the former configurations,  than the latter ones. But this   is at variance with the previous conclusion  indicating that the less stable model is described by the curve $d$.  On the other hand, however,  we observe from Figure \ref{fig:figure5}, that  the Tolman mass is not a monotonically increasing function of the radial coordinate, for the cases $c$ and $d$. The fact that the value of the Tolman mass, which we recall is a measure of the active gravitational mass, could be in some region, smaller than the value corresponding to an outer one, could be easily interpreted as a source of instability. In particular, it  suggests the possibility of a cracking (splitting) under perturbations \cite{cr1, p3}.

Thus, the distribution of the Tolman mass within the source, is characterized by two distinct physical properties, with opposite effects. One is the migration of the Tolman mass towards  the surface:   sharper it is, more stable is the model. This is a stabilization factor. The  second, is the change in the sign of  its spatial derivative: larger absolute values of negative derivatives enhance the instability of the  object.  This is a destabilizing factor. The superposition of these two factors leads to the specific degree of (un)stability of each model.
It is also worth noticing,  as shown in Fig. \ref{fig:figure6}, that curves $c$ and $d$ correspond to smaller  (absolute) values of   the anisotropy ($\Omega$)  in the outer regions, and  larger  ones in the inner regions.

\subsection{Case II}
For the case II we have run the program for an extensive  range of values of the  duplet $n, \alpha$, the final result of all these attempts being that none of the obtained models satisfy all the physical conditions (\ref{conditionsIII}). Since, in general, there exist anisotropic polytropes belonging to this case (see for example \cite{2p}), it is clear  that  physically meaningful models of this case, are ruled out by the conformally flat condition. However we do not know which (if any), is the specific rationale underlying this fact.

\section{conclusions}
We have resorted to the general method developed in Ref. \cite{2p} for the study of locally anisotropic polytropes, and we have applied it to the specific case of conformally flat  spheres. Our models are necessarily anisotropic (in pressure) and therefore are not continuously linked to isotropic polytropes. Another family of  anisotropic distributions, whose space of solutions is not simply connected  to the isotropic case, may be found in Ref. \cite{HPL}. 

Our solutions enabled us  to construct  models of highly compact spheres. The fact that such compact  spheres are in equilibrium even though the fluid in the neighborhood of the center is far from the relativistic regime ($\alpha \ll 1$), is explained by the specific anisotropy of pressure within the fluid distribution. This also explains  the change of sign in the gradients of $\psi_0$ and  the Tolman mass, and its peculiar distribution within the source. These two last effects, in turn, determine the (un)stability of the models, and  suggest, whenever the destabilizing factor prevails over the stabilization factor, the possible appearance of cracking in the fluid distribution, produced by perturbations of the fluid. Our models also seem to indicate  that while  the parameter $\alpha$ provides a good description about the regime of the fluid distribution in the isotropic case, it does no longer do so if pressure anisotropy is present.

Finally, although the specifics are out of the scope of this manuscript, it is worth mentioning that the models presented here could be useful in the discussion about the possible existence  of  super-Chandrasekhar white dwarfs.

Indeed, recent astrophysical evidence reveals the existence of overluminous type $Ia$  supernovae \cite{sup1, sup2, sup3, sup4, sup5, sup6}. One natural way to explain such observational data is to assume the existence of white dwarfs with masses above the Chandrasekhar limit, and which could act as the progenitors of such super  luminous type $Ia$ supernovae \cite{sup1, sup2, sup3, sup4, sup5, sup6}.
In order to assure the existence of such super-Chandrasekhar white dwarfs, two different mechanisms have been invoked so far.
On the one hand, it has been shown that  the Chandrasekar limit can be violated in the presence of strong magnetic fields  \cite{u1, u2, u3, u4, u5, u6, u7}.  On the other hand, specific charge distributions have been assumed, which also allows the existence of such super-Chandrasekhar white dwarfs \cite{chinos}.

At this point, four observations are in order:
\begin{itemize}
\item The magnetic field may be treated as an anisotropic fluid.  Indeed, it is a well-established fact  that  a magnetic field acting on a Fermi gas produces pressure anisotropy (see  Refs. \cite{ 23, 24, 25, 26} and references therein).
\item It should be emphasized that even though the spherical symmetry may be broken by the presence of a strong magnetic field, the assumption of such a symmetry may be considered as a good approximation  under a variety of circumstances (see \cite{u4, nov1, nov2}).
\item In both cases, the configurations are expected to be quite compact (large values of $y$).
\item  As we have seen, highly compact configurations may be obtained with the specific distribution of anisotropy created by the conformally flat condition. 
\end{itemize}

From the above comments, an obvious  question arises: can we obtain a distribution of local anisotropy (similar to the one of our models) by means of a physically meaningful, magnetic field and/or charge distribution, of the type one could expect to find   in a  white dwarf ? As important as it is, the answer to such a question lies beyond the scope of this work.

On the basis of all these comments, we envisage two possible directions to extend the results presented here:
\begin{enumerate}
\item Provide a physical picture for a source of the anisotropy, characteristic of the conformally flat models.
\item Apply the formalism which has been developed to detect the occurence of cracking \cite{cr1}, to the models considered here.
\end{enumerate}

\section*{Acknowledgments}
W.B. wishes to thank the Intercambio Cient\'\i fico Program, U.L.A., for financial support.
\end{document}